\documentclass[final,1p,times]{elsarticle} 
\usepackage{graphicx} 
\usepackage{amssymb} 
\usepackage{amsthm} 
\usepackage{lineno} 

\usepackage{epstopdf}

\newcommand{\be}{\begin{equation}}
\newcommand{\ee}{\end{equation}}
\newcommand{\ba}{\begin{eqnarray}}
\newcommand{\ea}{\end{eqnarray}}

\newcommand{\fig}{Fig.~}

\def\lsi{\raise0.3ex\hbox{$<$\kern-0.75em\raise-1.1ex\hbox{$\sim$}}}
\def\gsi{\raise0.3ex\hbox{$>$\kern-0.75em\raise-1.1ex\hbox{$\sim$}}}
\newcommand{\lsim}{\mathop{\lsi}}

\journal{Nuclear Physics A} 
\begin{document} 

\begin{frontmatter} 


\title{Towards the chiral critical surface of QCD}

\author{Owe Philipsen (in collaboration with Ph.de Forcrand)}

\address{Institut f\"ur Theoretische Physik, Westf\"alische Wilhelms-Universit\"at M\"unster,
48149 M\"unster, Germany}

\begin{abstract} 
The critical endpoint of the QCD phase diagram is usually
expected to belong to the chiral critical surface, i.e.~the surface of
second order transitions bounding the region of first order
chiral phase transitions for small quark masses in the $\{m_{u,d}, m_s,\mu\}$ parameter space.
For $\mu=0$, QCD with physical quark masses is known to be an analytic crossover,
requiring the region of chiral transitions to expand with $\mu$ for a critical endpoint to exist.
Instead, on coarse $N_t=4$ lattices, we find the area of chiral transitions to shrink with
$\mu$, which excludes a chiral critical point for QCD at moderate
chemical potentials $\mu_B < 500$ MeV.
First results on finer $N_t=6$ lattices indicate a curvature of the critical surface 
consistent with zero and unchanged conclusions.

\end{abstract} 

\end{frontmatter} 



\section{Introduction}\label{}

The QCD phase diagram has been the subject of intense research over the last ten years. 
Based on asymptotic freedom, one expects at least 
three different forms of nuclear matter: hadronic (low $\mu_B,T$), quark gluon plasma (high $T$) 
and colour-superconducting (high $\mu_B$, low $T$). Whether and where these regions are separated
by true phase transitions has to be determined by first principle calculations and experiments.
Since QCD is strongly coupled on scales of nuclear matter, Monte Carlo simulations
of lattice QCD are presently the only viable approach.

Unfortunately, the so-called sign problem prohibits straightforward
simulations at finite bary\-on density. There are several ways to circumvent this problem 
in an approximate way, all of them valid for $\mu/T\lsim1$ only \cite{oprev,csrev}. 
Within this range, all give
quantitatively agreeing results for, e.g., the calculation of $T_c(\mu)$ \cite{slavo}.
Because of the intricate and costly finite size scaling analyses involved, determining the order of
the transition, and hence the existence of a chiral critical point, is a much harder task.
Here we discuss the order of the finite temperature phase transition 
as obtained from lattice QCD simulations in the extended parameter space 
$\{m_{u,d},m_s,T,\mu\}$. 

\section{The chiral critical line at $\mu=0$}

\begin{figure}[t]
\begin{center}
\includegraphics[width=0.38\textwidth]{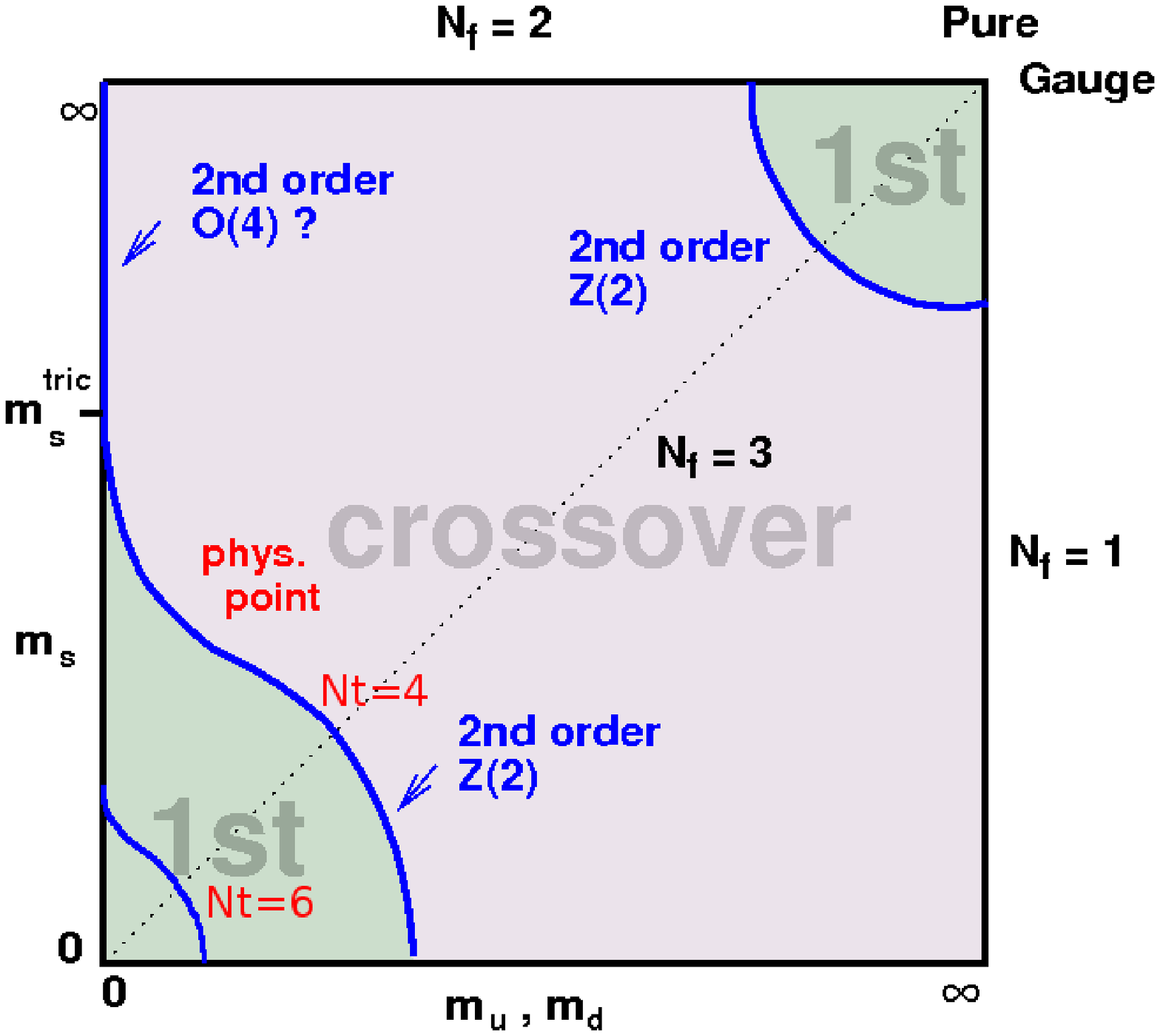}
\includegraphics[width=0.5\textwidth]{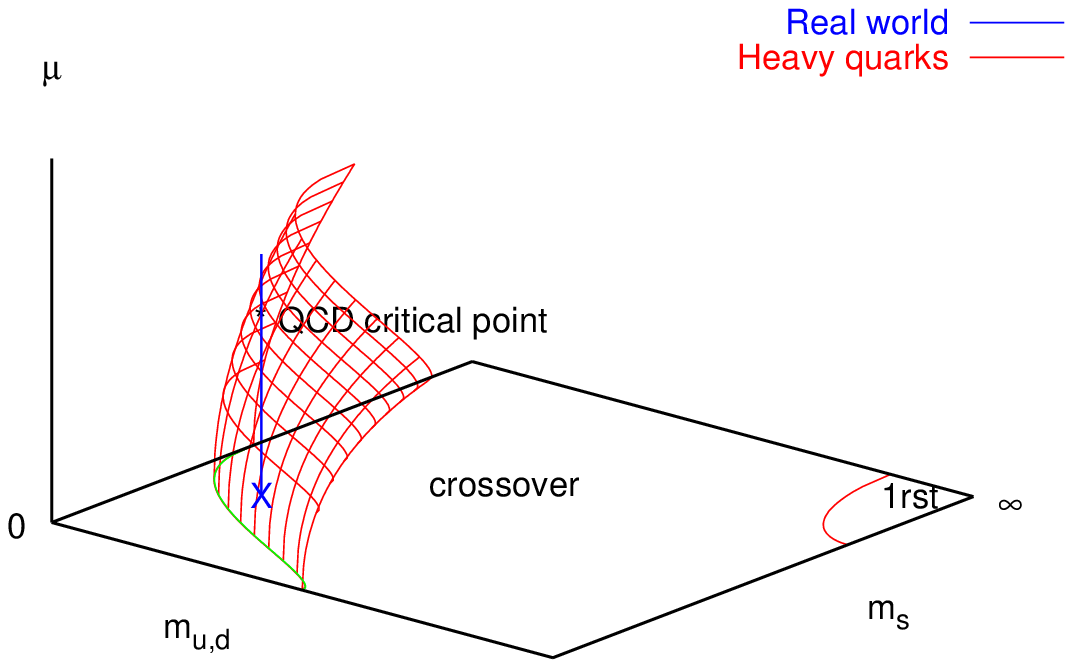}
\end{center}
\caption{\label{schem} Left: Schematic phase transition behaviour of $N_f=2+1$
QCD for different choices of quark masses at
$\mu=0$. On finer lattices, the chiral critical line moves towards smaller quark masses. 
Right: Critical surface swept by the chiral 
critical line as $\mu$ is turned on. Depending on its curvature, a QCD chiral critical
point is present or absent, cf.~\fig\ref{db4}. 
}
\end{figure}

The schematic situation is depicted in \fig\ref{schem}, beginning with $\mu=0$ (left).
In the limits of zero and infinite quark masses (lower left and upper 
right corners), order parameters corresponding to the breaking of a 
global symmetry can be defined, and one numerically finds first order phase
transitions at small and large quark masses at some finite
temperatures $T_c(m)$. On the other hand, one observes an analytic crossover at
intermediate quark masses, with second order boundary lines separating these
regions. Both lines have been shown to belong to the $Z(2)$ universality class
of the 3d Ising model \cite{kls,fp2,kim1}. Since the line on the lower left marks the boundary
of the quark mass region featuring a chiral phase transition, it is referred to as chiral critical line.
It has recently been mapped out on $N_t=4$ lattices \cite{fp3}.
A convenient observable is the Binder cumulant
$B_4(X) \equiv \langle (X - \langle X \rangle)^4 \rangle / \langle (X - \langle X \rangle)^2 \rangle^2$,
with $X \!=\! \bar\psi \psi$. At the second order transition, $B_4$ takes the value 1.604 dictated by the 
$3d$ Ising universality class.  
In agreement with expectations, the critical line steepens when approaching the chiral limit. Assuming 
the $N_f=2$ chiral transition to be in the $O(4)$ universality class implies a tricritical point on the $m_s$-axis, \fig\ref{schem} (left). The data are consistent with tricritical scaling \cite{derivs} of the critical line with $m_{u,d}$ and we estimate $m_s^{tric}\sim 2.8 T_c$. However, this value is extremely cut-off sensitive and likely smaller in the continuum, cf.~Sec.\ref{sec:nt6}.      

\section{The chiral critical surface}

\begin{figure}[t!]
\includegraphics[width=0.5\textwidth]{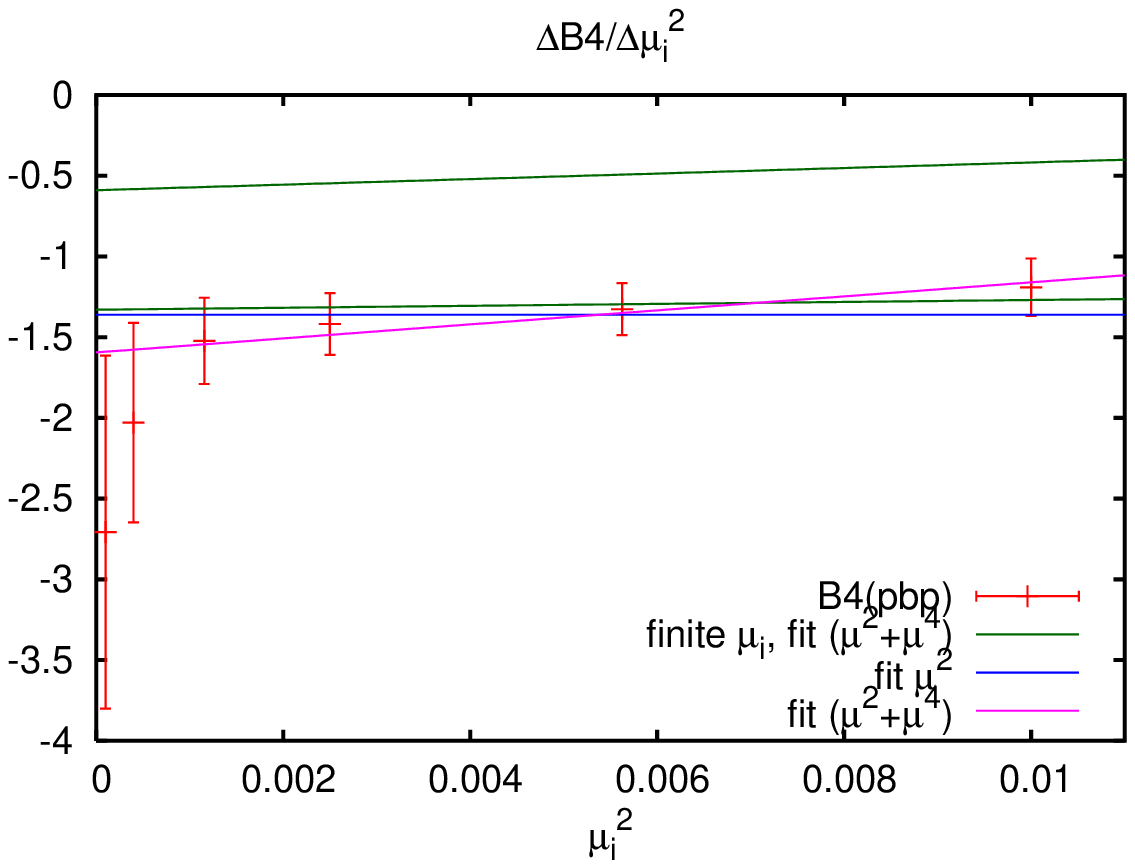}
\includegraphics[width=0.5\textwidth]{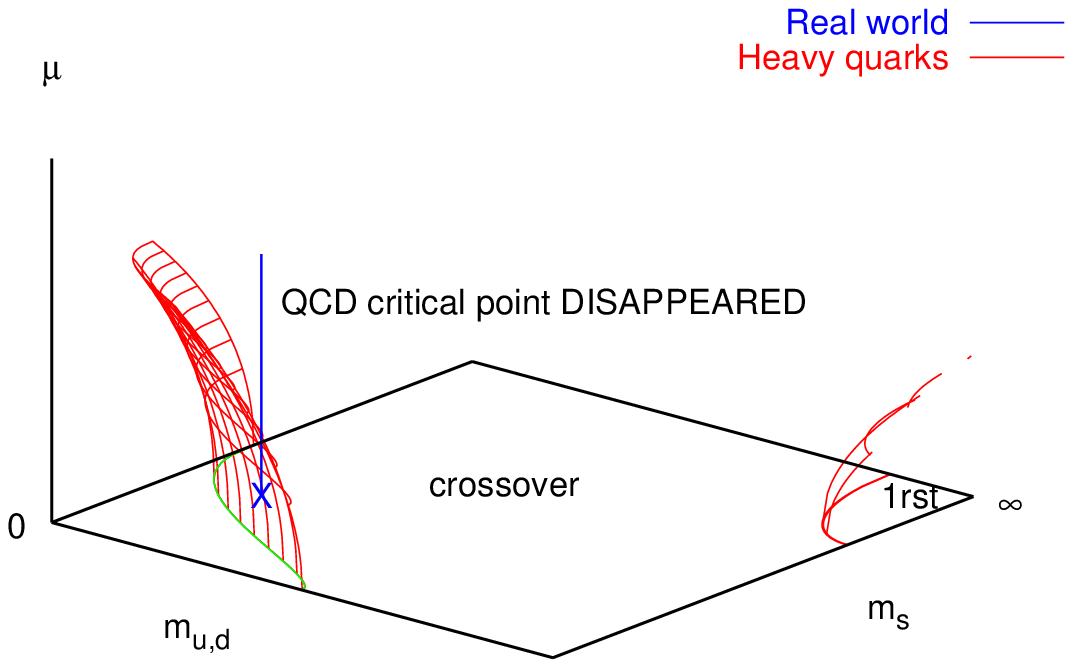}
\caption[]{$\mu^2$-dependence of the Binder cumulant on the chiral critical line for $N_f=3$ (left) \cite{fp4}. Thus, the scenario on the right is realised.
For heavy quarks the curvature has also been determined \cite{kim1} and the first order 
region shrinks with $\mu$.}
\label{db4}
\end{figure}

When a chemical potential is switched on, the chiral critical line sweeps out a surface, as shown
in \fig\ref{schem} (right). According to standard expectations \cite{derivs},
for small $m_{u,d}$, the critical line should 
continuously shift with $\mu$ to larger quark masses until it passes through the physical point at $\mu_E$, corresponding to the endpoint of the QCD phase diagram. 
This is depicted in \fig\ref{schem} (right), where the critical point is part of 
the chiral critical surface. However, it is also
possible for the chiral critical surface to bend towards smaller quark masses, cf.~\fig\ref{db4} (right),
in which case there would be no chiral critical point or phase transition 
at moderate densities. For definiteness, let us consider three degenerate quarks, 
represented by the diagonal in the quark mass plane.
The critical quark mass corresponding to the boundary point has an expansion
\be
\frac{m_c(\mu)}{m_c(0)}=1+\sum_{k=1}c_k \left(\frac{\mu}{\pi T}\right)^{2k}\,.
\ee
A strategy to learn about the chiral critical surface is now to tune the quark mass to $m_c(0)$ and evaluate
the leading coefficients of this expansion. In particular, the sign of $c_1$ will tell us which of the scenarios
is realised. 

The curvature of the critical surface in lattice units is directly related to the behaviour of the Binder cumulant via the chain rule,
\be
\frac{dam_c}{d(a\mu)^2}=-\frac{\partial B_4}{\partial (a\mu)^2}
\left(\frac{\partial B_4}{\partial am}\right)^{-1}\,.
\ee
While the second factor is sizeable and easy to evaluate, the $\mu$-dependence
of the cumulant is excessively weak and requires enormous statistics to extract. In order to guard
against systematic errors, this derivative has been evaluated in two independent ways.
One is to fit the corresponding Taylor series of $B_4$ in powers of $\mu/T$ to data generated at 
imaginary chemical potential \cite{fp3, fp4}, the other to compute the derivative directly and without
fitting via the finite difference quotient \cite{fp4}
\be
\frac{\partial B_4}{\partial (a\mu)^2}=\lim_{(a\mu)^2\rightarrow 0}\frac{B_4(a\mu)-B_4(0)}{(a\mu)^2}.
\ee 
Because the required shift in the couplings is very small,
it is adequate and safe to use the original Monte Carlo ensemble 
for $am^c_0,\mu=0$ and reweight the results by the standard 
Ferrenberg-Swendsen method. 
Moreover, by reweighting to imaginary $\mu$
the reweighting factors remain real positive and close to 1.
The results of these two procedures 
based on 20 and 5 million trajectories on $8^3\times 4$, respectively,  is shown in \fig\ref{db4} (left).
The error band represents the first coefficient from fits to imaginary $\mu$ data, while the
data points represent the finite difference quotient extrapolated to zero. Both results are consistent,
and the slope permits and extraction of the subleading $\mu^4$ coefficient. After continuum conversion the result for $N_f=3$ is $c_1=-3.3(3), c_2=-47(20)$  \cite{fp4}.
The same behaviour is found for non-degenerate quark masses. Tuning the strange quark 
mass to its physical value,  we calculated
$m^{u,d}_c(\mu)$ with $c_1= -39(8)$ and $c_2<0$, \fig\ref{nt6} (left).
Hence, on coarse $N_t=4$ lattices, the region of chiral phase transitions shrinks as a real chemical potential is turned on, and there is no chiral critical point for $\mu_B\lsim 500$ MeV, 
as in \fig\ref{db4} (right).
Note that one also observes a weakening of the phase transition with $\mu$ in the heavy quark case \cite{kim1}, in recent model studies of the light quark regime \cite{fuk,bow}, as well as a weakening of the transition with isospin chemical potential \cite{iso}.

\section{Towards the continuum, $N_t=6$}\label{sec:nt6}

\begin{figure}[t]
\begin{center}
\includegraphics[width=0.48\textwidth]{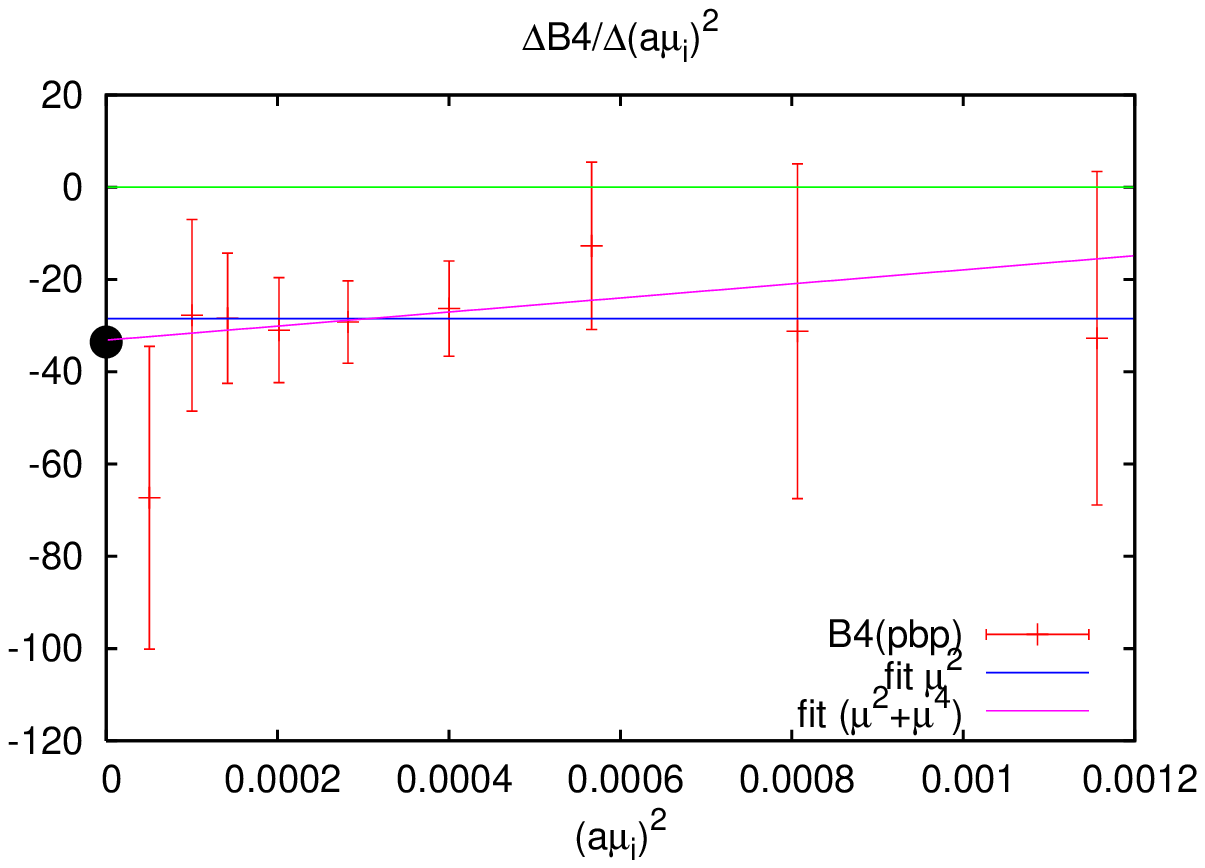}
\includegraphics[width=0.48\textwidth]{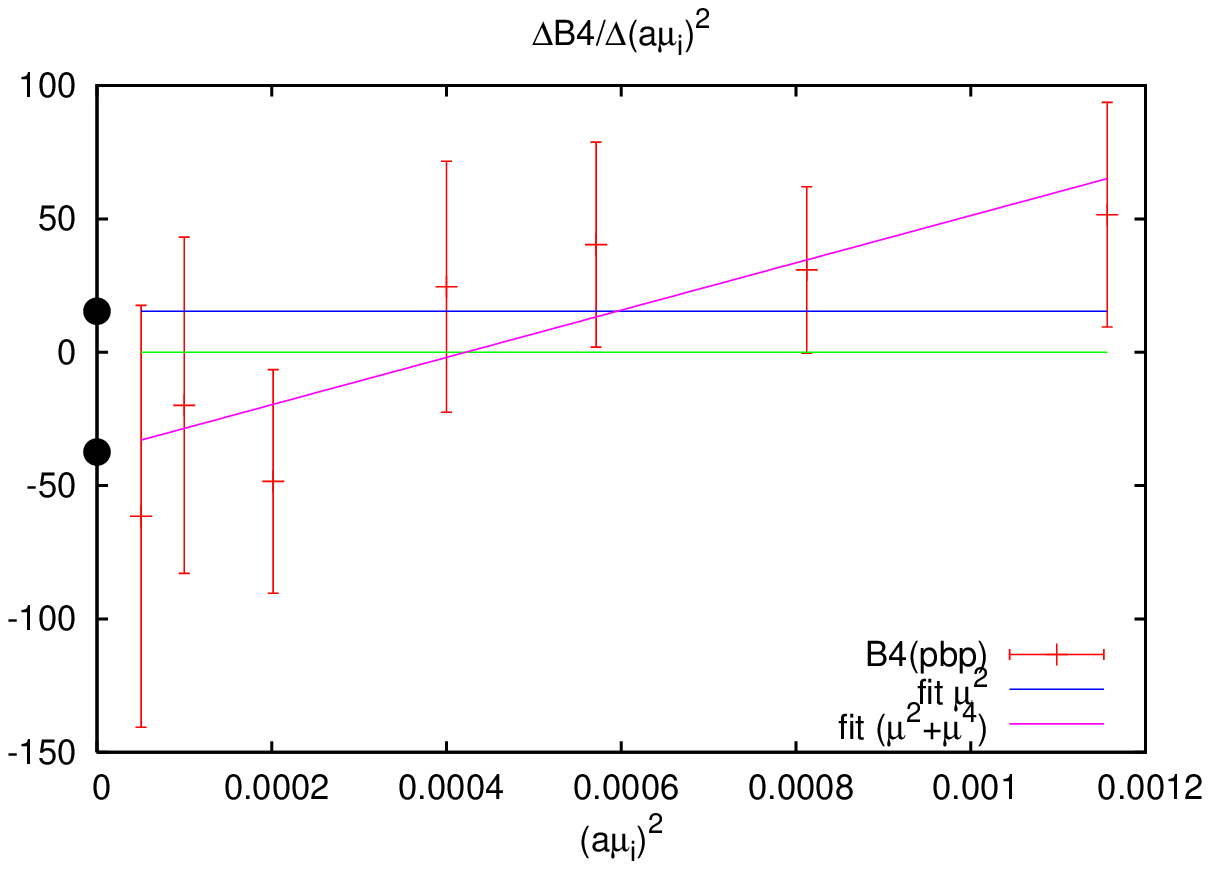}
\end{center}
\vspace*{-0.5cm}
\caption[]{Left:  $\mu^2$-dependence of the Binder cumulant on the chiral critical line for $N_f=2+1$ on $N_t=4$ with physical strange quark mass. 
Right: The same for $N_f=3,N_t=6$.}
\label{nt6}
\end{figure}

The largest uncertainty in these calculations by far is due to the coarse lattice spacing $a\sim 0.3$ fm
on $N_t=4$ lattices. First steps towards the continuum are currently being taken on $N_t=6, a\sim 0.2$ fm.  
At $\mu=0$, the chiral critical line is found to recede strongly with decreasing lattice 
spacing \cite{LAT07, fklat07}: for $N_f=3$, on the critical point $m_\pi(N_t=4)/m_\pi(N_t=6)\sim 1.8$.  
Thus, in the continuum the gap between the physical point and the chiral critical line
is much wider than on coarse lattices, as indicated in \fig\ref{schem} (left). 
Preliminary results for the curvature of the critical surface, \fig\ref{nt6} (right), result in 
$c_1=7(14),-17(18)$ for a LO,NLO extrapolation in $\mu^2$, respectively.
Thus the sign of the curvature is not yet constrained. But even if positive, 
its absolute size is too small to make up for the shift of the chiral critical line towards smaller quark masses, and one would again conclude for no chiral critical point below $\mu_B\lsim 500$ MeV
in this approximation. Higher order terms with large coefficients would be needed to
change this picture. 

However, on current lattices cut-off effects appear to be larger than finite density effects, hence 
definite conclusions for continuum physics cannot yet be drawn. A general finding is the 
steepness of the critical surface, making a possible critical endpoint etremely quark mass sensitive.
Furthermore, the shrinking of
the critical quark masses with diminishing lattice spacing makes it likely that the tricritical point moves to
$m_{u,d}=0,m_s^{tric}<m_s^{phys}$, in which case a possible critical endpoint might belong to an entirely different critical surface. 

%


\section*{Acknowledgments} 
This work is partially supported by the German BMBF, project 
No. 06MS254.

\end{document}